\documentclass{article}
\usepackage{spconf,amsmath,graphicx}
\usepackage{thmbox}
\usepackage{dsfont}
\usepackage{hyperref}
\usepackage{amsfonts}
\usepackage{amsmath}
\usepackage{amssymb}
\usepackage{rotating}
\usepackage{multirow}
\usepackage{array, makecell}
\usepackage{bm}
\usepackage{stmaryrd}
\usepackage{arydshln}
\usepackage{colortbl}
\newtheorem[L]{thm}{Theorem}[section]
\newtheorem{lemma}[thm]{Lemma}
\newtheorem{proposition}[thm]{Proposition}
\usepackage{mathtools}

\usepackage{fancyhdr}
\title{Towards a unified view of unsupervised non-local methods for image denoising: the NL-Ridge approach}
\name{Sébastien Herbreteau \quad Charles Kervrann\thanks{Thanks to  Bpifrance agency for funding through the LiChIE contract.}}
\address{SERPICO Project-Team, Inria Centre Rennes - Bretagne Atlantique, \\
UMR144 CNRS Institut Curie, PSL Research University, Sorbonne Université, France}
\begin{document}
%
\maketitle

\begin{abstract}
We propose a unified view of unsupervised non-local
methods for image denoising that linearily combine noisy image patches. The best methods, established in different modeling and estimation frameworks, are two-step algorithms. Leveraging Stein’s unbiased risk estimate (SURE) for the first step and the "internal adaptation", a concept borrowed from deep learning theory, for the second one, we show that our NL-Ridge approach enables to reconcile several patch aggregation methods for image denoising. In the second step, our closed-form aggregation weights are computed through multivariate Ridge regressions. Experiments on artificially noisy images demonstrate that NL-Ridge may outperform well established state-of-the-art unsupervised denoisers such as BM3D and NL-Bayes, as well as  recent unsupervised deep learning methods, while being simpler conceptually. 
\end{abstract}
\section{Introduction}
\label{sec:intro}

From anisotropic diffusion to recently deep learning-based methods, image denoising is probably one of the most important topic in image processing. Among them, the N(on)L(ocal)-means algorithm \cite{nlmeans} that exploits the self-similarity assumption and information redundancy may be considered as a milestone in image denoising. It is assumed that, in a natural image, a patch rarely appears alone: almost perfect copies can be found in its surroundings.
The NL-means algorithm \cite{nlmeans} then amounts to computing, for each pixel, an average of its neighboring noisy pixels, weighted by the degree of similarity of patches they belong to. This inspired a lot of methods afterward that manage several groups of similar noisy patches \cite{BM3D}, \cite{nlbayes}, \cite{WNNM}, \cite{PLR}. In the final step, patches are re-positionned to their initial locations and a final averaging is performed at a given position as each pixel has been denoised multiple times. To exploit patches more collaboratively, frequency-based methods \cite{BM3D} or low-rank assumptions \cite{WNNM}, \cite{PLR} were proposed  and led to state-of-the-art performances for more than a decade.  

In this paper, our contribution is twofold. First, we propose a unified view, well grounded in Stein’s unbiased risk estimation theory, to reinterpret and reconcile previous state-of-the-art non-local methods. In our estimation framework, no prior model for the distribution on patches is required. Second, the resulting two-step NL-Ridge algorithm, based on closed-form expressions, may outperform BM3D \cite{BM3D} and NL-Bayes \cite{nlbayes}, as well as several unsupervised deep-learning methods \cite{S2S} \cite{N2S} \cite{DIP} 
on artificially noisy images with additive white Gaussian noise.

\section{NL-Ridge for image denoising}
\label{section2}

\subsection{A two-step approach applied to patches}

Leveraging the self-similarity assumption, our objective is to design a local denoiser $f_\Theta$, where $\Theta$ is a set of parameters, that operates on groups of similar noisy patches extracted from a noisy image and combines them collaboratively. Formally, let $X \in \mathbb{R}^{n \times m}$ be a matrix gathering $m$ clean similar patches of size $\sqrt{n} \times \sqrt{n}$ and $Y \in \mathbb{R}^{n \times m}$ its noisy version. The matrices $X$  and $Y$ are generally referred to similarity matrices in the literature. One wants to estimate the matrix $X$ from $Y = X + W$,  where $W \in \mathbb{R}^{n \times m}$ is a matrix whose elements $W_{i,j} \sim  \mathcal{N}(0, \sigma^2)$  (Gaussian noise) are assumed to be independent along each row and each column, respectively. Our local denoiser is of the form:
\begin{equation}
f_{\Theta} : Y \mapsto Y \Theta
\label{local_denoiser}
\end{equation}

\noindent where $\Theta \in \mathbb{R}^{m \times m}$. The function $f_{\Theta}$ aims here at denoising the $m$ noisy similar patches contained in $Y$  all at once. To this end, each noisy patch is processed by a linear combination of its most similar patches whose weights are gathered in the matrix $\Theta$. Aggregating similar patches  via a linear combination  has already been  exploited in the past \cite{nlmeans} \cite{PEWA} \cite{OWF}. However, the originality of our method lies in the way of computing the weights $\Theta$, which significantly boosts performance.

The optimal local denoiser $f_\Theta$  is found by minimizing the quadratric risk defined as:
\begin{equation}
R_\Theta(X) = \mathbb{E}\|f_\Theta(Y) - X\|^2_F
\label{riskDef}
\end{equation}
where $\| \cdot \|^2_F$  denotes the Frobenius norm. In other words, we look for the Minimum-Mean-Squared-Error (MMSE) estimator among the family of functions $(f_\Theta)_{\Theta \in \mathbb{R}^{m \times m}}$. The optimal estimator $f_{\Theta^\ast}$ minimizes the risk, \textit{i.e.}
\begin{equation}
    \Theta^\ast = \arg \min_{\Theta} \;  R_{\Theta}(X).
\label{eq2}
\end{equation}
Unfortunately, $\Theta^\ast$ requires the knowledge of $X$  which is unknown. The good news is that the risk $R_{\Theta}(X)$  can be approximated through the following two-step algorithm:

\begin{itemize}
    \item In the first step, an approximation of $\Theta^\ast$ is computed for each group of similar patches, through the use of an unbiased estimate of $R_{\Theta}(X)$, Stein's unbiased risk estimate (SURE) \cite{SURE}. After reprojection \cite{Aggreg} of all denoised patches, a first denoised image $\hat{I}_1$ is obtained.
    \item In the second step, $\hat{I}_1$ is improved with a second estimation of $\Theta^\ast$ which is found thanks to the technique of "internal adaptation" described in \cite{LIDIA}.
\end{itemize}

\subsection{Step 1: Stein's unbiased risk estimate}

Preliminarily, let us consider  the two following propositions.

\begin{proposition} Let $Y = X + W$  where $Y, X, W \in \mathbb{R}^{n \times m}$ and $W_{i,j} \sim {\cal N}(0, \sigma^2)$  are independent along each row. An unbiased estimate of the risk $R_\Theta(X)  = \mathbb{E}\|f_\Theta(Y) - X\|^2_F$ is Stein's unbiased risk estimate (SURE):
$$\operatorname{SURE}_{\Theta}(Y) = \| Y \Theta - Y \|_F^2 + 2n\sigma^2  \operatorname{tr}(\Theta) - nm\sigma^2$$
where $\operatorname{tr}$ denotes the trace operator.
\label{proposition1NLRidge}
\end{proposition}

\begin{proposition} Let $Y \in \mathbb{R}^{n \times m}$ and $\Theta \in \mathbb{R}^{m \times m}$.
$$\hat{\Theta}_1  = \arg \min_{\Theta} \; \operatorname{SURE}_{\Theta}(Y) 
= I_m - n \sigma^2 (Y^\top Y)^{-1} $$ 
where $I_m$ is the $m \times m$ identity matrix and $\top$ denotes the transpose operator.
\label{proposition2NLRidge}
\end{proposition}

Our objective is to get an approximation of $\Theta^\ast$ from (\ref{eq2}). Proposition \ref{proposition1NLRidge} gives an unbiased estimate of the risk $R_{\Theta}(X)$ that does not depend on $X$, but only on the observations $Y$. A common idea that has been previously exploited in image denoising (\textit{e.g.} see \cite{surenlmeans}, \cite{surelet}, \cite{sureGM}) is to use this estimate as a surrogate for minimizing the risk $R_{\Theta}(X)$ in (\ref{eq2}) which is inaccessible. 
Based on Proposition \ref{proposition2NLRidge}, we get a first estimation: 
\begin{equation}
\hat{\Theta}_{1} = I_m - n \sigma^2 (Y^\top Y)^{-1}.
\label{theta1}
\end{equation}

Note that $\hat{\Theta}_1$ is close to $\Theta^{\ast}$ as long as the variance of SURE is low. A rule of thumbs used in \cite{surelet} states that the number of parameters must not be "too large" compared to the number of data in order for the variance of SURE to remain small. In our case, this suggests that $m < n$.

\subsection{Step 2: Internal adaptation}

\begin{proposition} 
The quadratic risk $R_{\Theta}(X)$ defined in (\ref{riskDef}) is:
$$ \mathbb{E}\|f_\Theta(X+W) - X\|^2_F = \| X \Theta - X \|_F^2 + n\sigma^2 \| \Theta \|_F^2$$
which is minimal for:
$$\Theta^{\ast} = ( X^\top X + n\sigma^2 I_m)^{-1} X^\top X $$
(solution of the multivariate Ridge regression). 
\label{proposition3NLRidge}
\end{proposition}

At the end of the first step, we get a first denoised image $\hat{I}_{1}$ that will serve as a pilot in the second step. Once again, we focus on the solution of (\ref{eq2}) to denoise locally similar patches. As $X$ and $\hat{X}_{1}$, the corresponding group of similar patches in $\hat{I}_{1}$, are supposed to be close, the "internal adaptation" procedure \cite{LIDIA} consists in solving (\ref{eq2}) by substituting $\hat{X}_{1}$ for $X$. From Proposition \ref{proposition3NLRidge}, we get the following closed-form solution:
\begin{equation}
\hat{\Theta}_{2} = ( \hat{X}_{1}^\top \hat{X}_{1} + n\sigma^2 I_m)^{-1} \hat{X}_{1}^\top \hat{X}_{1}
\label{theta2}
\end{equation}

\noindent which is nothing else than the solution of a multivariate Ridge regression. Interestingly, in practice, this second estimate of $\Theta^\ast$ produces a significant boost in terms of denoising performance compared to $\hat{\Theta}_{1}$. The second step can be iterated but we did not notice improvements in our experiments.

\subsection{Weighted average reprojection}

After the denoising of a group of similar patches, each denoised patch is repositionned at its right location in the image. As several pixels are denoised multiple times, a final step of  aggregation, or reprojection \cite{Aggreg}, is necessary to produce a final denoised image $\hat{I}_1$ or $\hat{I}_2$. With inspiration from \cite{Aggreg}, each pixel belonging to column $j$ of $Y$ is assigned, after denoising, the weight $w_j  = 1 / ( \|\Theta_{\cdot,j} \|_2^{2})$. Those weights are at the end pixel-wise normalized such that the sum of all weights associated to a same pixel equals one.

\section{A unified view of non-local denoisers}
\label{section3}

In NL-Ridge, the local denoiser $f_\Theta$ is arbitrarily of the form given by (\ref{local_denoiser}) involving the weighted aggregation of similar patches with closed-form weights given in (\ref{theta1}) and (\ref{theta2}).   In this section, we show that NL-Ridge can serve to interpret two popular state-of-the-art non-local methods (NL-Bayes \cite{nlbayes} and BM3D \cite{BM3D}), which were originally designed with two very different modeling and estimation frameworks. It amounts actually to considering two particular families $(f_\Theta)$ of local denoisers. 

\subsection{Analysis of NL-Bayes algorithm}

The NL-Bayes \cite{nlbayes} algorithm has been established in the Bayesian setting and the resulting maximum a posteriori estimator is computed with a two-step procedure as NL-Ridge. As starting point, let us consider the following local denoiser:
\begin{equation}
f_{\Theta, \beta} : Y \mapsto  \Theta Y + \beta u^\top
\end{equation}
\noindent where $\Theta \in \mathbb{R}^{n \times n}$, $\beta \in \mathbb{R}^{n}$, and $u$ denotes a $m$-dimensional vector with all entries equal to one.

 \noindent \textbullet \; \textbf{Step 1:} Stein's unbiased estimate of the quadratic risk $R_{\Theta, \beta}(X)$ reaches its minimum for:
\begin{equation}
    \hat{\Theta}_{1} = (C_Y - \sigma^2 I_n)C_Y^{-1} \quad \text{and} \quad  \hat{\beta}_{1}  = (I_n - \hat{\Theta}_{1} ) \mu_Y
\end{equation}
\noindent where $\mu$ and $C$ denote the empirical mean and covariance matrix of a group of patches. Interestingly, $f_{\hat{\Theta}_{1}, \hat{\beta}_{1}}(Y)$ is the expression given in \cite{nlbayes} (first step), which is actually derived from the prior distribution of patches assumed to be Gaussian. Furthermore, our framework provides guidance on the choice of the parameters $n$ and $m$. Indeed, SURE is helpful provided that its variance remains small which is achieved if $n < m$ (the number of parameters must not be “too large” compared to the number of data). This result suggests that NL-Bayes is expected to be performant if small patches are used, as confirmed in the experiments in \cite{nlbayesImp}.

\noindent \textbullet \; \textbf{Step 2:} Interpreting the second step in \cite{nlbayes} as an "internal adaptation" step, the updated parameters become: 
\begin{equation}
    \hat{\Theta}_{2} = C_{\hat{X}_1}(C_{\hat{X}_1} + \sigma^2 I_n)^{-1} \quad \text{and} \quad  \hat{\beta}_{2}  = (I_n - \hat{\Theta}_{2})\mu_{\hat{X}_1}
\end{equation}
and $f_{\hat{\Theta}_{2}, \hat{\beta}_{2}}(Y)$ corresponds to the original second-step expression in \cite{nlbayes}. 

\subsection{Analysis of BM3D algorithm}

 BM3D \cite{BM3D} is probably the most popular non-local method for image denoising. It assumes a locally sparse representation of images in a transform domain. A two step algorithm  was described in \cite{BM3D} to achieve state-of-the-art results for several years. By using the generic  NL-Ridge  formulation, we consider the following family:
\begin{equation}
f_{\Theta} : Y \mapsto  P^{-1} (\Theta \odot (P Y Q)) Q^{-1} 
\end{equation}
\noindent where $\Theta \in \mathbb{R}^{n \times m}$ and $\odot$ denotes the Hadamard product. $P \in \mathbb{R}^{n \times n}$ and $Q \in \mathbb{R}^{m \times m}$ are two orthogonal matrices that model a separable 3D-transform (typically a 2D and 1D \textit{Discrete Cosine Transform}, respectively).

\noindent \textbullet \; \textbf{Step 1:} The minimization of SURE yields:  
\begin{equation}
    \hat{\Theta}_{1a} = U - \frac{\sigma^2}{(PYQ)^2} 
\end{equation}

\noindent where $U \in \mathbb{R}^{n \times m}$ whose elements equal to one and where division and square are  element-wise operators. Unfortunately, $f_{\hat{\Theta}_{1a}}(Y)$ does not provide very satisfying denoising results. This result is actually expected as the number of parameters equals the size of data ($nm$), and makes SURE weakly performant.  To overcome this difficulty, we can force the elements of $\Theta$ to be either $0$ or $1$. Minimizing SURE with this contraint yields to:
\begin{equation}
    \hat{\Theta}_{1b} = \mathds{1}(|PYQ| \geq \sqrt{2}\sigma). 
\end{equation}
\noindent where $\mathds{1}(x \geq \tau) = 1$ if $x \geq \tau$ and $0$ otherwise. Our denoiser $f_{\Theta}$ is then a hard thresholding estimator as in BM3D: the coefficients of the transform domain (\textit{i.e.} the elements of the matrix $PYQ$) below  $\sqrt{2}\sigma$, in absolute value, are cancelled before applying the inverse 3D-transform. This result suggests that the threshold should be linearly dependent on $\sigma$ but also independent on orthogonal transforms $P$ and $Q$. In \cite{BM3D}, a threshold  of $2.7\sigma$ was carefully chosen in Step 1, which is approximatively twice the SURE-prescribed threshold.

\noindent \textbullet \; \textbf{Step 2:} the "internal adaptation" yields the same expression as the Wiener filtering step in BM3D: 
\begin{equation}
    \hat{\Theta}_{2} = \frac{(P\hat{X}Q)^2}{\sigma^2 + (P\hat{X}Q)^2}
\end{equation}

In conclusion, we have shown that BM3D \cite{BM3D} and NL-Bayes \cite{nlbayes} can be interpreted within NL-Ridge framework, enabling to set the size of the patches and to potentially relax the need to specify the prior distribution of patches.

\section{Experimental results}
\label{section4}

In this section, we compare the performance of our NL-Ridge
method with state-of-the-art methods, including related learning-based methods \cite{dncnn} \cite{ffdnet} \cite{LIDIA} \cite{S2S} \cite{N2S} \cite{DIP} applied to standard gray images artificially corrupted with additive white Gaussian noise with zero mean and variance $\sigma^2$. Performances of NL-Ridge and other methods are assessed in terms of PSNR values. We want to emphasize that NL-Ridge is as fast as BM3D \cite{BM3D} and NL-Bayes \cite{nlbayes} as they share the same paradigm. It has been implemented in Python with Pytorch, enabling it to run on GPU unlike its traditional counterparts. The code can be downloaded at: \href{https://github.com/sherbret/NLRidge/}{https://github.com/sherbret/NL-Ridge/}.

\addtolength{\tabcolsep}{-4pt} 
\begin{figure}[ht]
\centering
\renewcommand{\arraystretch}{0.5}
\begin{tabular}{cc}
 \includegraphics[scale=0.23]{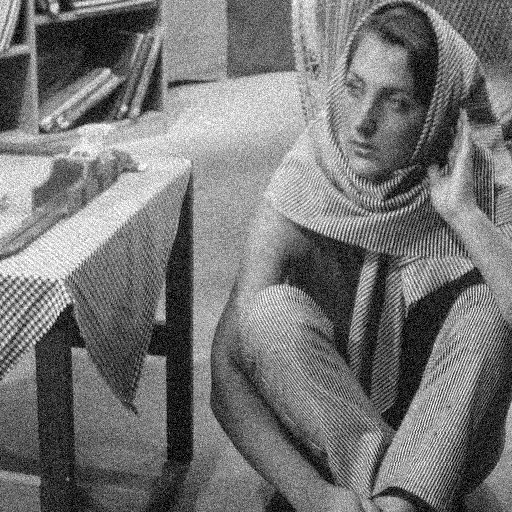} &
\includegraphics[scale=0.23]{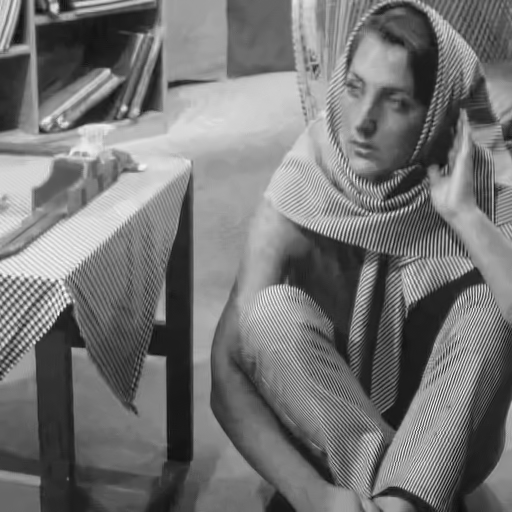} \\
\footnotesize Noisy / 22.09 dB  &   \footnotesize BM3D / 31.72 dB  \\
\includegraphics[scale=0.23]{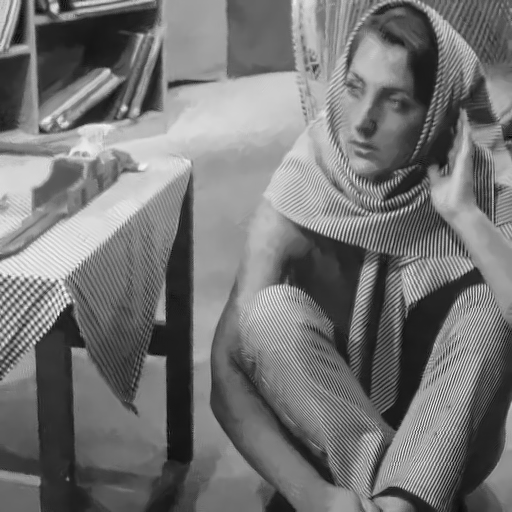} &\includegraphics[scale=0.23]{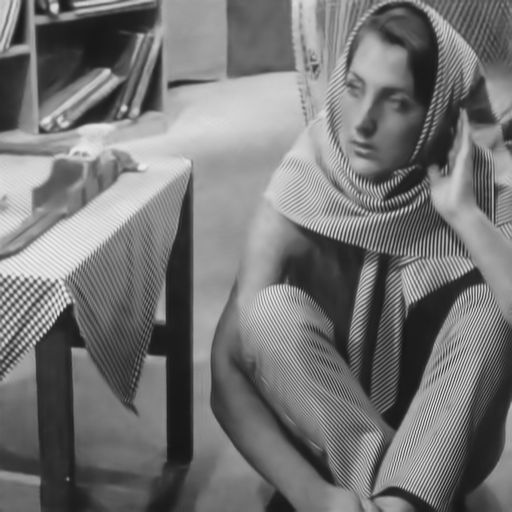} \\
\footnotesize NL-Bayes / 31.54 dB &\footnotesize Self2Self / 31.62 dB   \\
\includegraphics[scale=0.23]{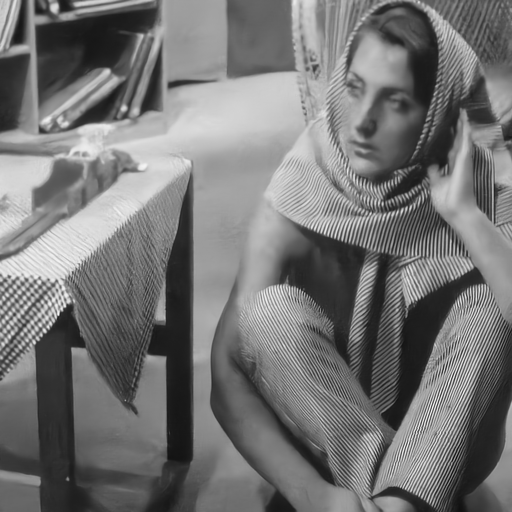} &
\includegraphics[scale=0.23]{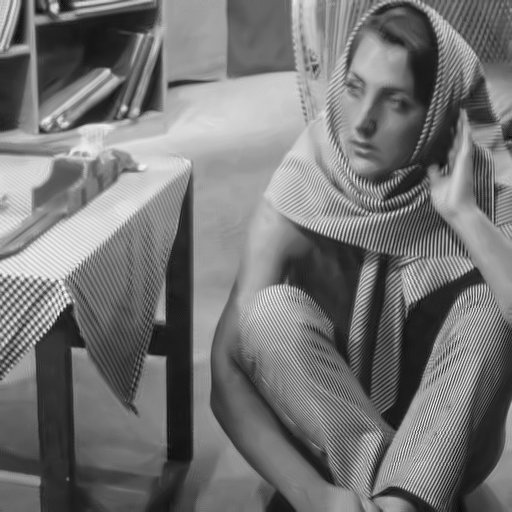} \\
\footnotesize DnCNN / 31.06 dB  & \footnotesize NL-Ridge (ours) / \textbf{32.06} dB \\
\end{tabular}
\caption{Denoising results (in PSNR) on \textit{Barbara} corrupted with additive white Gaussian noise ($\sigma = 20$).}
\label{photo}
\end{figure}
\addtolength{\tabcolsep}{4pt}

\subsection{Setting of algorithm parameters}

For the sake of computational efficiency, the search for groups of similar patches across the image is restricted to a small local window of size $w \times w$ (in our experiments $w=45$). More precisely, for a reference patch $y_j$, one looks for its $m-1$ more similar patches of size $\sqrt{n} \times \sqrt{n}$ in the local window in order to form the similarity matrix $Y$ of size $n \times m$. We used the  conventional $L_2$ distance to form groups of similar patches. Considering iteratively each overlapping patch of the image as reference patch is also computationally demanding. To speed up the algorithm, NL-Ridge is performed on a subsampled position grid.  We used a moving step $\delta = 4$ from one reference patch to its neighbors both horizontally and vertically. 

\begin{table}[ht]
\caption{Recommended patch size $n$ and patch number $m$ for the Step 1 and Step 2 versus noise standard deviation $\sigma$.}
\centering
\begin{tabular}{*{5}{|c}|}
  \hline
   $\sigma$ & $n_{1}$ & $n_{2}$ & $m_{1}$ & $m_{2}$ \\\hline\hline
 $\textcolor{white}{1}0< \sigma \leq 15$ & $7\times7$ & $7\times7$ & 18 & 55\\\hline
 $15 < \sigma \leq 35$ & $9\times9$ & $9\times9$ & 18 & 90\\\hline
 $35 < \sigma \leq 50$ & $11\times11$ & $9\times9$ & 20 & 120\\\hline
\end{tabular}
\label{optimalParams}
\end{table}

Finally, the choice of the parameters $n$ and $m$ depend on the noise level. Experimentally, bigger patches have to be considered for higher noise levels as well as a higher quantity of patches for the second step. An empirical analysis leads to choose the parameters reported in Table \ref{optimalParams}.

\addtolength{\tabcolsep}{-3pt}
\begin{table}[ht]
\centering
  \caption{The average PSNR (dB) results of different methods on various datasets corrupted with Gaussian noise ($\sigma=15$ and $25$). Best performance among each category is in bold.}\label{set12tab}
  
  \begin{tabular}{|c @{\hspace{0.01cm}} ccccc |}
  \hline
  &&  Methods & Set12 & BSD68 & Urban100 \\\hline\hline
 
  && Noisy &  24.61 / 20.17 &  24.61 / 20.17 &  24.61 / 20.17 \\\hline
 \multirow{6}{*}{\begin{sideways} \scriptsize \textit{Unsupersived} \end{sideways}} & \multirow{3}{*}{\begin{sideways} \scriptsize \textit{Trad.} \end{sideways}} & BM3D &  32.37 / 29.97 &  31.07 / 28.57 &  32.35  / 29.70 \\
 && NL-Bayes  &  32.25 / 29.88 &  31.16 / \textbf{28.70} &  31.96 / 29.34 \\
 && \textbf{NL-Ridge}  &  \textbf{32.46} / 30.00  &  \textbf{31.20} / 28.67 &  \textbf{32.53} / \textbf{29.90} \\\cdashline{2-3}
 & \multirow{3}{*}{\begin{sideways} \scriptsize \textit{DL} \end{sideways}} & DIP  &   30.12  / 27.54 &  28.83 / 26.59 &  -  / -  \\
 
 && Noise2Self    &  31.01 / 28.64  &  29.46 / 27.72 &  - / - \\
 && Self2Self  &  32.07 / \textbf{30.02} &  30.62  / 28.60 &  - / - \\\hline
 
 \multirow{3}{*}{\begin{sideways} \scriptsize \textit{Super-} \end{sideways}} & \multirow{3}{*}{\begin{sideways} \scriptsize \textit{vised} \end{sideways}} & DnCNN &  \textbf{32.86} / \textbf{30.44} &  \textbf{31.73} / \textbf{29.23} &  32.68 / 29.97  \\
 && FFDnet  &  32.75 / 30.43 & 31.63 / 29.19 &  32.43 / 29.92 \\
 && LIDIA  &  32.85 / 30.41 &  31.62 / 29.11 &  \textbf{32.80} / \textbf{30.12} \\\hline
\end{tabular}
\label{resultsPSNR}
\end{table}
\addtolength{\tabcolsep}{0.5pt}

\subsection{Results on test datasets}

We tested the denoising performance of our method on
three well-known datasets: Set12, BSD68 \cite{berkeley} and Urban100 \cite{urban}. A comparision with state-of-the-art algorithms is reported in Table \ref{resultsPSNR}. For a fair comparison, algorithms are divided into two categories: unsupervised methods (either traditional or deep learning-based) and supervised deep learning-based ones that require a training phase beforehand on an external dataset. We used the implementations provided by the authors for all algorithms. As for Noise2Self \cite{N2S}, only the single-image extension was considered.

NL-Ridge, exclusively based on weighted aggregation of noisy patches, performs surprisingly at least as well as its traditional counterparts \cite{BM3D} \cite{nlbayes}. It is particularly efficient on Urban100 dataset which contains abundant structural patterns and textures, achieving comparable performances with FFDnet \cite{ffdnet}, a popular supervised network composed of hundreds of thousands of parameters.

Figure \ref{photo} illustrates the visual results of different methods.  NL-Ridge  is very competitive with respect to well-established methods such as BM3D \cite{BM3D}. The self-similarity assumption is particularly useful to recover subtle details such as the stripes on the \textit{Barbara} image that are better reconstructed than DnCNN \cite{dncnn}.

\section{Conclusion}

We presented a unified view to reconcile state-of-the-art unsupervised non-local denoisers through the minimization of a risk from a family of estimators, exploiting Stein's unbiased risk estimate on the one hand and the "internal adaptation" on the other. We derive NL-Ridge algorithm, which leverages local linear combinations of noisy similar patches. Our experimental results show that NL-Ridge compares favourably with its state-of-the-art counterparts, including recent unsupervised deep learning methods which are much more computationally demanding. 

\bibliographystyle{IEEEbib}
\bibliography{bibli}

\onecolumn
\section{Supplementary material}
\subsection{Useful lemmas}

\begin{lemma} 
Let $A, B \in \mathbb{R}^{n \times m}$, $\Theta \in \mathbb{R}^{m \times m}$ and $W \in \mathbb{R}^{n \times m}$ a random matrix where the components  $W_{i, j} \sim \mathcal{N}(0, \sigma^2)$ are independent along each row.
$$\mathbb{E}\|(A+W) \Theta -B\|^2_F = \|  A \Theta - B \|_F^2 + n\sigma^2 \|   \Theta  \|_F^2$$
\label{lemma1}
\end{lemma}

\begin{proof} $\| (A + W) \Theta - B \|_F^2 = \| A \Theta - B  \|_F^2  + \| W \Theta \|_F^2  + 2 \langle A \Theta - B,  W \Theta  \rangle_F $

\noindent Now computing the expected value yields:
$$\mathbb{E}\langle A \Theta - B,  W \Theta  \rangle_F = 0$$ and, as $W_{i,j}$ are independent along each row, $$\begin{aligned}\mathbb{E}\| W \Theta \|_F^2 = \mathbb{E} \left( \sum_{i=1}^{n} \sum_{j=1}^{m} \left( \sum_{k=1}^{m} W_{i,k} \Theta_{k,j} \right)^2 \right) 
=  \sum_{i=1}^{n} \sum_{j=1}^{m} \mathbb{E}  \left( \sum_{k=1}^{m} W_{i,k} \Theta_{k,j} \right)^2  = \sum_{i=1}^{n} \sum_{j=1}^{m}  \sum_{k=1}^{m} \sigma^2 \Theta_{k,j}^2 = n\sigma^2  \| \Theta \|_F^2
\end{aligned}$$
\end{proof}

\begin{lemma} 
Let $A, B \in \mathbb{R}^{n \times m}$, $\Theta \in \mathbb{R}^{m \times m}$ and $\lambda \in \mathbb{R}$. If $A^\top A$ is invertible:
$$\arg \min_{\Theta} \|  A \Theta - B \|_F^2 + 2\lambda \operatorname{tr}(\Theta) = (A^\top A)^{-1} (A^T B - \lambda I_m)$$
\label{lemma2}
\end{lemma}

\begin{proof}
Let $g: \Theta  \mapsto \| A \Theta - B \|_F^2 + 2 \lambda  \operatorname{tr}(\Theta)$.

\noindent  We have $\displaystyle g(\Theta) = \sum_{j=1}^{m} \| A \Theta_{\cdot, j} - B_{\cdot, j} \|_2^2 + 2 \lambda  \Theta_{j,j} = \sum_{j=1}^{m} h_j(\Theta_{\cdot, j})$ \; with \; $ \displaystyle h_j : \theta \in \mathbb{R}^{m} \mapsto \| A \theta - B_{\cdot, j} \|_2^2 + 2 \lambda  \theta_{j}$. 

\noindent Then, $\nabla h_j(\theta) = 2A^\top(A \theta - B_{\cdot, j}) + 2 \lambda e_j = 0 \; \Leftrightarrow \; \theta = (A^\top A)^{-1} (A^\top B_{\cdot, j}  - \lambda  e_j).$

\noindent Hence,
$\displaystyle  \arg \min_{\Theta_{\cdot, j}} h_j(\Theta_{\cdot, j})  = (A^\top A)^{-1} (A^\top B_{\cdot, j}  - \lambda  e_j)$ \; and \; $ \displaystyle \arg \min_{\Theta} g(\Theta)  = (A^\top A)^{-1} (A^\top B  - \lambda I_m).$
\end{proof}

\begin{lemma} 
Let $A, B \in \mathbb{R}^{n \times m}$, $\Theta \in \mathbb{R}^{m \times m}$ and $\lambda > 0$.
$$\arg \min_{\Theta} \|  A \Theta - B \|_F^2 + \lambda \| \Theta \|_F^2 = (A^\top A + \lambda I_m)^{-1} A^T B $$
\label{lemma3}
\end{lemma}

\begin{proof}
Solution of a multivariate Ridge regression.
\end{proof}

\begin{lemma} 
Let $A \in \mathbb{R}^{n \times m}$, $\beta \in \mathbb{R}^n$ and $v \in \mathbb{R}^m$ with $v \neq 0$.
$$\arg \min_{\beta} \|  A - \beta v^\top \|_F^2 = \frac{Av}{\| v \|_2^2} $$
\label{lemma4}
\end{lemma}

\begin{proof} $\displaystyle \|  A - \beta v^\top \|_F^2  = \sum_{i=1}^{n} \|  A_{i, \cdot} - \beta_i v \|_2^2 = \sum_{i=1}^{n} \|  A_{i, \cdot} \|_2^2 + \beta_i^2 \|  v \|_2^2 - 2 \beta_i \langle A_{i, \cdot}, v  \rangle  $

\noindent $\displaystyle \frac{\partial}{\partial \beta_j} \|  A - \beta v^\top \|_F^2  = 2 \beta_j \| v \|_2^2 - 2 \langle A_{j, \cdot}, v  \rangle = 0  \; \Leftrightarrow \; \beta_j = \frac{\langle A_{j, \cdot}, v  \rangle}{\| v \|_2^2}$, hence, $\displaystyle \arg \min_{\beta} \|  A - \beta v^\top \|_F^2 = \frac{Av}{\| v \|_2^2}$.
\end{proof}

\subsection{Proofs for NL-Ridge}
\noindent \underline{\textbf{Proposition 2.1:}} For $n=1$, all components of $Y$ are independent and Stein's unbiased risk estimate is given by \cite{SURE}:
$$\operatorname{SURE}_{\Theta}(Y) = - m\sigma^2 + \| f_{\Theta}(Y) - Y \|_F^2 + 2\sigma^2 \operatorname{div} f_{\Theta}(Y)$$
\noindent with $\displaystyle \operatorname{div} f_{\Theta}(Y) = \sum_{j=1}^{m} \frac{\partial f_{\Theta j}}{\partial y_{j}}(Y) = \sum_{j=1}^{m} \frac{\partial }{\partial y_{j}} Y \Theta_{\cdot,j} = \sum_{j=1}^{m} \frac{\partial }{\partial y_{j}} \sum_{k=1}^{m} Y_{1,k} \Theta_{k,j} =  \sum_{j=1}^{m} \Theta_{j,j}= \operatorname{tr}(\Theta)$.

\noindent For $n \geq 1$, $\begin{aligned} \mathbb{E}\| f_{\Theta}(Y) - X \|_F^2     
= \sum_{i=1}^{n} \mathbb{E} \| Y_{i, \cdot} \Theta - X_{i, \cdot} \|_F^2  
= \sum_{i=1}^{n} \mathbb{E} ( \operatorname{SURE}_{\Theta}(Y_{i, \cdot})) 
= \mathbb{E} ( \| Y \Theta - Y \|_F^2  + 2n\sigma^2  \operatorname{tr}(\Theta) - nm\sigma^2 ).\end{aligned}$

\noindent \underline{\textbf{Proposition 2.2:}} Using Lemma \ref{lemma2},
$$\arg \min_{\Theta} \; \| Y \Theta - Y \|_F^2 + 2n\sigma^2\operatorname{tr}(\Theta) = (Y^\top Y)^{-1} (Y^\top Y - n \sigma^2 I_m) = I_m - n \sigma^2 (Y^\top Y)^{-1}$$ 

\noindent \underline{\textbf{Proposition 2.3:}} Using Lemma \ref{lemma1} and \ref{lemma3},
$$ R_X(\Theta) = \mathbb{E} \| f_{\Theta}(X+W) - X \|_F^2 = \mathbb{E} \| (X + W) \Theta - X \|_F^2 = \|  X \Theta - X \|_F^2 + n\sigma^2 \|   \Theta  \|_F^2 $$
which is minimal for:
$$\Theta^{\ast} = ( X^\top X + n\sigma^2 I_m)^{-1} X^\top X $$

\subsection{Proofs for NL-Bayes}

The local denoiser in NL-Bayes is of the form:
$$f_{\Theta, \beta} : Y \mapsto  \Theta Y + \beta u^\top$$
\noindent where $\Theta \in \mathbb{R}^{n \times n}$, $\beta \in \mathbb{R}^{n}$ and $u \in \mathbb{R}^{m}$ a vector composed of ones.

\begin{proposition} 
Let $Y = X + W$  where $Y, X, W \in \mathbb{R}^{n \times m}$ and $W_{i,j} \sim {\cal N}(0, \sigma^2)$  are independent along each column. An unbiased estimate of the risk $R_{\Theta, \beta}(X)  = \mathbb{E}\|f_{\Theta, \beta}(Y) - X\|^2_F$ is Stein's unbiased risk estimate (SURE):
$$\operatorname{SURE}_{\Theta, \beta}(Y) = \|  \Theta Y   + \beta u^\top  - Y \|_F^2 + 2 m \sigma^2 \operatorname{tr}(\Theta) - nm\sigma^2$$
\end{proposition}

\begin{proof}
For $m=1$, all components of $Y$ are independent and Stein's unbiased risk estimate is given by \cite{SURE}:
$$\operatorname{SURE}_{\Theta, \beta}(Y) = - n\sigma^2 + \| f_{\Theta, \beta}(Y) - Y \|_F^2 + 2\sigma^2 \operatorname{div} f_{\Theta, \beta}(Y)$$
with $\displaystyle \operatorname{div} f_{\Theta, \beta}(Y) = \sum_{i=1}^{n} \frac{\partial f_{\Theta, \beta i}}{\partial y_{i}}(Y) = \sum_{i=1}^{n} \frac{\partial }{\partial y_{i}} \sum_{k=1}^{n} \Theta_{i,k} Y_{k,1} =  \sum_{i=1}^{n} \Theta_{i,i}= \operatorname{tr}(\Theta)$.

\noindent $\begin{aligned} \displaystyle  \text{For } m \geq 1, \; \mathbb{E}\| f_{\Theta, \beta}(Y) - X \|_F^2 
= \sum_{j=1}^{m} \mathbb{E} \| \Theta Y_{\cdot, j} + u_j \beta   - X_{\cdot, j} \|_F^2  
&= \sum_{j=1}^{m} \mathbb{E} ( \operatorname{SURE}_{\Theta, \beta}(Y_{ \cdot, j}))  \\
&= \mathbb{E} ( -nm\sigma^2 +  \|  \Theta Y   + \beta u^\top  - Y \|_F^2 + 2 m \sigma^2 \operatorname{tr}(\Theta) ) \end{aligned}.$
\end{proof}

\begin{proposition}
Let $Y \in \mathbb{R}^{n \times m}$, $\Theta \in \mathbb{R}^{n \times n}$ and $\beta \in \mathbb{R}^{n}$.
$$\hat{\Theta}_1, \hat{\beta}_{1} = \arg \min_{\Theta, \beta} \operatorname{SURE}_{\Theta, \beta}(Y) = (C_Y - \sigma^2 I_n) C_Y^{-1},  (I_n - \hat{\Theta}_{1} ) \mu_Y $$
\end{proposition}

\begin{proof}
Using Lemma \ref{lemma4}, for $\Theta$ fixed, $\operatorname{SURE}_{\Theta, \beta}(Y)$ is minimal for $\displaystyle \beta = -\frac{(\Theta Y - Y ) u}{\|  u \|_2^2}.$ Injecting it in the expression of SURE yields:
$\displaystyle \|   Y^{\top} \Theta^{\top}    -\frac{u  u^\top}{\|  u \|_2^2}  (\Theta Y - Y )^\top  - Y^{\top} \|_F^2 + 2 m \sigma^2 \operatorname{tr}(\Theta^{\top}) = \|   Z \Theta^{\top} - Z \|_F^2 + 2 m \sigma^2 \operatorname{tr}(\Theta^{\top})$ with $Z = Y^{\top} -\frac{u  u^\top}{\|  u \|_2^2} Y^{\top}$, which is minimal, using lemma \ref{lemma2}, for $\displaystyle \hat{\Theta}^{\top}_{1} =  (Z^\top Z)^{-1} (Z^\top Z - m \sigma^2 I_n).$ Now, for $u$ composed of ones, $Z^\top Z = Y Y^\top -\frac{1}{m}  Yuu^\top Y^\top = m C_Y$. Finally, $\displaystyle \hat{\Theta}_1 =   (C_Y- \sigma^2 I_n) C_Y^{-1}$ and 
$\displaystyle \hat{\beta}_{1} = -\frac{(\hat{\Theta}_1 Y - Y ) u}{\|  u \|_2^2} = \frac{Yu}{m} - \hat{\Theta}_1 \frac{Yu}{m} = (I_n - \hat{\Theta}_{1} ) \mu_Y$.
\end{proof}

\begin{proposition} 
The quadratic risk $R_{\Theta, \beta}(X)$ has the explicit form:
$$R_{\Theta, \beta}(X) = \mathbb{E} \| f_{\Theta,  \beta}(X+W) - X \|_F^2 = \|   \Theta X  +  \beta u^\top - X \|_F^2 + m\sigma^2 \| \Theta \|_F^2 $$
which is minimal for:
$$\hat{\Theta}_{2} = C_{X}(C_{X} + \sigma^2 I_n)^{-1} \quad \text{and} \quad  \hat{\beta}_{2}  = (I_n - \hat{\Theta}_{2})\mu_{X}$$
\end{proposition}

\begin{proof}
Using Lemma \ref{lemma1},
$$\begin{aligned} R_{\Theta, \beta}(X) = \mathbb{E} \| \Theta (X+W)  + \beta u^\top - X\|^2_F &= \mathbb{E} \|  (X^\top +W^\top) \Theta^\top  + u \beta^\top - X^\top \|^2_F \\
&= \|  X^\top \Theta^\top  + u \beta^\top - X^\top \|_F^2 + m\sigma^2 \|   \Theta^\top  \|_F^2 \\
&= \|   \Theta X  +  \beta u^\top - X \|_F^2 + m\sigma^2 \| \Theta \|_F^2
\end{aligned}$$

\noindent For $\Theta$ fixed and using Lemma \ref{lemma4}, it is minimized for $\displaystyle \beta = -\frac{(\Theta X - X)u}{ \| u \|_2^2}$.

\noindent Injecting it in the expression of the risk:
$\displaystyle \|  X^\top \Theta^\top  + \frac{u u^\top}{\| u \|_2^2} (X^\top-  X^\top \Theta^\top) - X^\top \|_F^2 + m\sigma^2 \|   \Theta^\top  \|_F^2
= \|  Z  \Theta^\top - Z \|_F^2 + m\sigma^2 \|   \Theta^\top  \|_F^2$  with $\displaystyle Z = X^\top  - \frac{u u^\top}{\| u \|_2^2} X^\top$. This quantity is minimal, using Lemma \ref{lemma3}, for $\displaystyle \hat{\Theta}^\top_{2} = (Z^\top Z + m\sigma^2 I_n)^{-1} Z^\top Z$.

\noindent Now, for $u$ composed of ones, $Z^\top Z = XX^\top - \frac{1}{m} X u u^\top X^\top = m C_X$.

\noindent Finally, $\displaystyle \hat{\Theta}_2 =  C_X (C_X+\sigma^2 I_n)^{-1} $ \; and \; $ \displaystyle \hat{\beta}_{2} = -\frac{(\hat{\Theta}_2 X - X ) u}{\|  u \|_2^2} = \frac{Xu}{m} - \hat{\Theta}_2 \frac{Xu}{m} = (I_n - \hat{\Theta}_{2} ) \mu_X.$
\end{proof}

\subsection{Proofs for BM3D}
The local denoiser in BM3D is of the form:
$$f_{\Theta} : Y \mapsto  P^{-1} (\Theta \odot (P Y Q)) Q^{-1} $$
\noindent where $\Theta \in \mathbb{R}^{n \times m}$ and $P \in \mathbb{R}^{n \times n}$ and $Q \in \mathbb{R}^{m \times m}$ are two orthogonal matrices: $PP^\top = I_n$ and $QQ^\top = I_m$. 

\begin{proposition} 
Let $Y = X + W$  where $Y, X, W \in \mathbb{R}^{n \times m}$ and $W_{i,j} \sim {\cal N}(0, \sigma^2)$  are independent. An unbiased estimate of the risk $R_{\Theta}(X)  = \mathbb{E} \|  f_{\Theta}(Y) - X \|_F^2$ is Stein's unbiased risk estimate (SURE):
$$\operatorname{SURE}_{\Theta}(Y) = \|  P^{-1} (\Theta \odot (P Y Q)) Q^{-1} - Y \|_F^2  +2\sigma^2 \langle \Theta, U \rangle_F -nm\sigma^2$$
\noindent where $U \in \mathbb{R}^{n \times m}$ is a matrix composed of ones and $\langle \cdot, \cdot \rangle_F$ denotes the Frobenius inner product.
\label{propBM3D}
\end{proposition}

\begin{proof}

\noindent By development of the squared Frobenius norm, $ \; \displaystyle \|  f_{\Theta}(Y) - Y \|_F^2 = \|  f_{\Theta}(Y) - X \|_F^2 + \| W  \|_F^2 - 2 \langle f_{\Theta}(Y) - X, W \rangle_F.$ As $P^{-1} = P^\top$ and $Q^{-1} = Q^\top$:
$$\begin{aligned}\langle f_{\Theta}(Y), W \rangle_F = \langle P^{-1} (\Theta \odot (P Y Q)) Q^{-1}, W \rangle_F 
&= \langle \Theta \odot (P Y Q), PWQ \rangle_F  \\
&= \langle \Theta \odot (P X Q), PWQ  \rangle_F + \langle \Theta \odot (P W Q), PWQ  \rangle_F \end{aligned}$$

\noindent Now computing the expected value for each term yields:
$$ \mathbb{E} \langle  \Theta \odot (PXQ) , PWQ \rangle_F = 0, \quad \mathbb{E} \| W  \|_F^2 = nm\sigma^2, \quad \mathbb{E} \langle X, W \rangle_F = 0 \quad \text{and} \quad \mathbb{E} \langle  \Theta \odot (PWQ)  , PWQ \rangle_F  = \sigma^2 \langle U, \Theta \rangle_F.$$

\noindent Indeed, as the $W_{i,j}$ are independent and $P$ and $Q$ are orthogonal matrices:
$$\begin{aligned}
\mathbb{E} [\Theta_{i,j} (PWQ)_{i,j}^2] = \Theta_{i,j} \mathbb{E} [ (PWQ)_{i,j}^2] = \Theta_{i,j} \left(\underbrace{\mathbb{E} [ (PWQ)_{i,j}]^2}_{=0} + \mathbb{V} [ (PWQ)_{i,j}] \right) &= \Theta_{i,j} \mathbb{V} \left( \sum_{k=1}^{m} \left(\sum_{l=1}^{n} P_{i, l} W_{l, k}  \right) Q_{k, j}  \right) \\
&= \Theta_{i,j}  \sum_{k=1}^{m}  Q_{k, j}^2 \mathbb{V} \left(\sum_{l=1}^{n} P_{i, l} W_{l, k}  \right)   \\
&= \Theta_{i,j}  \sum_{k=1}^{m}  Q_{k, j}^2  \sum_{l=1}^{n} P_{i, l}^2 \mathbb{V} \left(W_{l, k}  \right)   \\
&= \Theta_{i,j}  \sum_{k=1}^{m}  Q_{k, j}^2  \sum_{l=1}^{n} P_{i, l}^2 \sigma^2     \\
&=\sigma^2 \Theta_{i,j}
\end{aligned}$$

\noindent Finally, we get $\;  \mathbb{E} \|  f_{\Theta}(Y) - X \|_F^2 = \mathbb{E} \left[  \|  f_{\Theta}(Y) - Y \|_F^2  +2\sigma^2 \langle U, \Theta \rangle_F -nm\sigma^2 \right].$
\end{proof}

\begin{proposition}
Let $Y \in \mathbb{R}^{n \times m}$, $\Theta \in \mathbb{R}^{n \times m}$ and $U \in \mathbb{R}^{n \times m}$ a matrix composed of ones.
$$\hat{\Theta}_{1a} = \arg \min_{\Theta} \; \operatorname{SURE}_{\Theta}(Y) = U - \frac{\sigma^2}{(PYQ)^2}$$

$$\hat{\Theta}_{1b} = \arg \min_{\substack{\Theta \\ s.t. \; \Theta_{i, j} \in \{0, 1\}}} \; \operatorname{SURE}_{\Theta}(Y) = \mathds{1}(|PYQ| \geq \sqrt{2} \sigma)$$
\end{proposition} 

\begin{proof} As $P$ and $Q$ are orthogonal matrices:
$$\|  f_{\Theta}(Y) - Y \|_F^2 + 2\sigma^2 \langle U, \Theta \rangle_F = \| \Theta \odot (PYQ)  - PYQ \|_F^2  +2\sigma^2 \langle U, \Theta \rangle_F = \sum_{i=1}^{n} \sum_{j=1}^{n} ( (PYQ)_{i,j} \Theta_{i,j}  - (PYQ)_{i,j})^2  + 2\sigma^2 \Theta_{i,j} $$

\noindent Let $\alpha \in \mathbb{R}^\ast$. The minimum of $x \in \mathbb{R} \mapsto (\alpha x - \alpha)^2 + 2\sigma^2 x$ is obtained for $\displaystyle x^{a}_{min} = 1 - \frac{\sigma^2}{\alpha^2}$. Hence, $$ \hat{\Theta}_{1a} = \arg \min_{\Theta} \|  f_{\Theta}(Y) - Y \|_F^2 + 2\sigma^2 \langle U, \Theta \rangle_F = U - \frac{\sigma^2}{(PYQ)^2}.$$

\noindent The minimum of $x \in \{0, 1\} \mapsto (\alpha x - \alpha)^2 + 2\sigma^2 x$ is obtained for $\displaystyle x^b_{min} = \mathds{1}(|\alpha| \geq \sqrt{2}\sigma)$. Hence, $$ \hat{\Theta}_{1b} = \mathds{1}(|PYQ| \geq \sqrt{2} \sigma).$$
\end{proof}

\begin{proposition} 
The quadratic risk $R_{\Theta}(X)$ has the explicit form:
$$R_{\Theta}(X) = \mathbb{E} \| f_{\Theta}(X+W) - X \|_F^2 = \|   \Theta \odot PXQ  - PXQ \|_F^2 + \sigma^2 \| \Theta \|_F^2 $$
which is minimal for:
$$\hat{\Theta}_{2} = \frac{(PXQ)^2}{\sigma^2 + (PXQ)^2}$$
\end{proposition}

\begin{proof}
As $P$ and $Q$ are orthogonal matrices:
$$ \begin{aligned} \|  f_{\Theta}(X+W) - X \|_F^2  &=  \|  P^{-1} (\Theta \odot (P(X + W) Q))Q^{-1} - X \|_F^2 \\
&=  \|  \Theta \odot (PXQ) + \Theta \odot  (PWQ) - PXQ \|_F^2 \\
&=  \|  \Theta \odot (PXQ) - PXQ  \|_F^2 + \| \Theta \odot  (PWQ) \|_F^2  + 2 \langle \Theta \odot (PXQ) - PXQ, \Theta \odot  (PWQ) \rangle_F \\
\end{aligned}$$

\noindent Now computing the expected value for each term yields:
$$\mathbb{E} \langle \Theta \odot (PXQ) - PXQ, \Theta \odot  (P W Q) \rangle_F = 0 $$
\noindent and
$$ \begin{aligned}
\mathbb{E} \| \Theta \odot  (PWQ) \|_F^2 = \sum_{i=1}^{n} \sum_{j=1}^{m} \mathbb{E} [(\Theta_{i,j} (PWQ)_{i,j})^2] &= \sum_{i=1}^{n} \sum_{j=1}^{m}  \mathbb{V} [\Theta_{i,j} (PWQ)_{i,j}] + \underbrace{\mathbb{E} [\Theta_{i,j} (PWQ)_{i,j}]^2}_{=0}  \\
&= \sum_{i=1}^{n} \sum_{j=1}^{m} \Theta_{i,j}^2 \underbrace{\mathbb{V} [ (PWQ)_{i,j}]}_{=\sigma^2} \quad \textit{(see proof from proposition \ref{propBM3D}}) \\
&= \sigma^2  \|  \Theta \|_F^2 \\
\end{aligned}$$

\noindent Finally, \; $ \displaystyle \mathbb{E} \|  f_{\Theta}(X+W) - X \|_F^2 =  \|  \Theta \odot PXQ  - PXQ \|_F^2 + \sigma^2 \| \Theta \|_F^2 = \sum_{i=1}^{n} \sum_{j=1}^{m} (PXQ)_{i,j}^2 (\Theta_{i,j} - 1)^2  + \sigma^2  \Theta_{i,j}^2$.

\noindent Let $\alpha \in \mathbb{R}$. The minimum of $x \mapsto \alpha^2(x - 1)^2 + \sigma^2 x^2$ is obtained for $\displaystyle x = \frac{\alpha^2}{\sigma^2 + \alpha^2}$. Finally,
$$  \arg \min_{\Theta} \;  \|  \Theta \odot PXQ  - PXQ \|_F^2 + \sigma^2 \| \Theta \|_F^2 = \frac{(PXQ)^2}{\sigma^2 + (PXQ)^2}.$$

\end{proof}

\end{document}